NEW JERSEY INSTITUTE OF TECHNOLOGY

# Improvement of inter-protocol fairness for BBR congestion control using machine learning

Asif Kunwar[1], Khushi Jain[2], Vaishnavi Mhaske[3], Sai Kartik Thatikonda[4]

[1]*Professor, Department of Computer Science, NJIT, Newark NJ, USA*

[2,3,4]*Student, Department of Computer Science, NJIT, Newark NJ, USA*

***Abstract*: Google's BBR (Bottleneck Bandwidth and Round-trip Propagation Time) approach is used to enhance internet network transmission. It is particularly intended to efficiently handle enormous amounts of data. Traditional TCP (Transmission Control Protocol) algorithms confront the most difficulty in calculating the proper quantity of data to send in order to prevent congestion and bottlenecks. This wastes bandwidth and causes network delays. BBR addresses this issue by adaptively assessing the available bandwidth (also known as bottleneck bandwidth) along the network channel and calculating the round-trip time (RTT) between the sender and receiver. Although when several flows compete for bandwidth, BBR may supply more bandwidth to one flow at the expense of another, resulting in unequal resource distribution. This paper proposes to integrate machine learning with BBR to enhance fairness in resource allocation. This novel strategy can improve bandwidth allocation and provide a more equal distribution of resources among competing flows by using historical BBR data to train an ML model. Further we also implemented a classifier model that is graphic neural network in the congestion control method.**

## 1. Introduction

The expansion of varied apps and services has resulted in an extraordinary spike in data traffic on the internet in recent years. The requirement for effective congestion management systems has grown in importance to maintain equitable resource allocation and optimal network performance. Because of its capacity to dynamically adapt to changing network circumstances and produce high throughput, the Bottleneck Bandwidth and Round-trip propagation time (BBR) congestion management algorithm has emerged as a potential alternative. While BBR performs admirably in a variety of circumstances, questions have been raised about its inter-protocol fairness, particularly when coexisting with other congestion control methods. Interactions between BBR and older protocols in diverse network settings might result in an unjust distribution of network resources, resulting in unsatisfactory user experiences and hurting overall network efficiency[1].

```
For an MPTCP connection, judge shared-bottleneck sets
  in ProbeBW and ProbeRTT states and count the number
  n_i of subflows in each shared-bottleneck set i
for each shared bottleneck set S_i do
    Initialize α_r = 1/n_i,
    for each subflow r do
        if the subflow r in the shared-bottleneck set then
            RTT' = get_current_maxRTT();
            BDP_r = max{x_r(t)} × RTT'      t ∈ [T − W, T];
            α_r = max BtlBW_r / Σ_{r∈S} BtlBW_r;
            G_r ∈ [1.25, 0.75, α_r, α_r, α_r, α_r, α_r, α_r];
            if Infight packets ≤ BDP_r then
                R_r = G_r × max{x_r(t)};
            else if Infight packets > BDP_r then
                R_r = 0;
            end

    end
end
for each of other independent subflows, r , which
  doesn't belong to any shared-bottleneck set do
    r acts as a single-path TCP-BBR flow;
end
```

**Figure 1(a) Bottleneck Fairness Considered Coupled Congestion Control Algorithm**

The goal of this study is to use machine learning approaches to address the problem of inter-protocol fairness in BBR congestion control. Machine learning has proven to be effective in optimizing complicated systems, and its application to network protocols has the potential to increase fairness, reduce congestion, and, ultimately, improve customer pleasure.

**Our study aims to reimplement the paper by conducting a survey of existing models to evaluate their efficiency and real-time responsiveness to identify the model that is most effective and quickly adapts to changing network environments. In the context of our research, an integral**

**component of our future endeavors involves the implementation of Neural Network Classifier. We planned to gather diverse datasets containing features like throughput, latency etc. and preprocess the data to split it into training and testing data. Following that, we delve into machine learning principles and their application in the context of network optimization. Finally, we give experimental findings and assessments that show how our strategy improves inter-protocol fairness and network performance.**

**We desire to provide helpful insights and approaches to the larger area of congestion control through our study, enabling a fairer and more efficient network environment in the face of developing communication technologies.**

## 2. Methodology

To improve network performance, the study used a Linux system to implement the TCP BBR congestion control method. The procedure comprised system reboots, compatibility checks, and kernel package updates. Important actions included making sure data was backed up, confirming BBR activation, and reviewing prerequisites. The approach prioritized correctness and readability, stressing compliance with Linux distribution documentation. To aid in the optimization of network protocols, a standardized method for installing TCP BBR was to be established [2].

**Figure 2(a) Implementation of TCP BBR in Linux Figure 2**

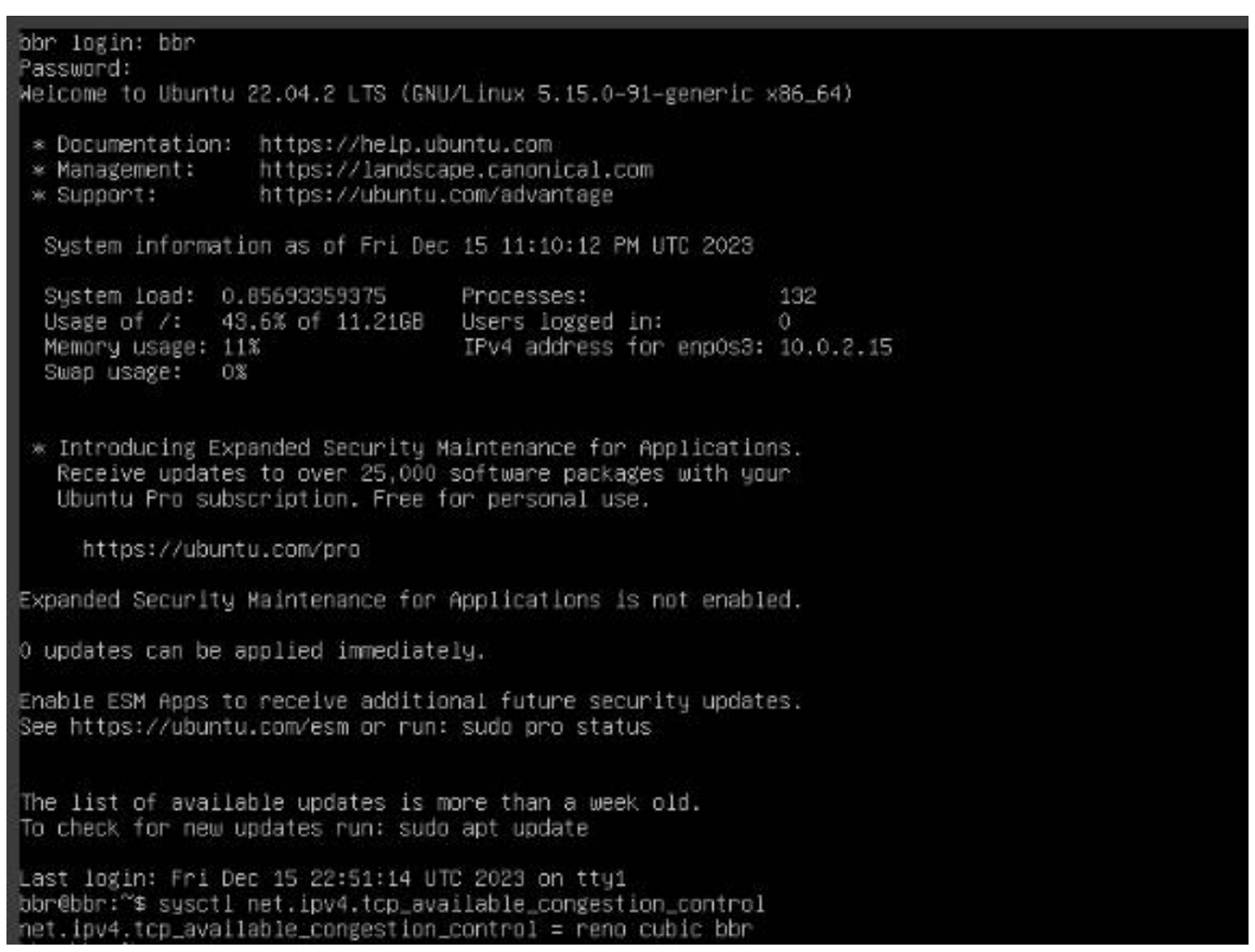

The Transmission Control Protocol (TCP) was used in the research approach to create a connection and collect network packets. Using packet capture tools, establishing a TCP connection, and setting network interfaces were crucial tasks. Clarity and reproducibility were given a priority, with a focus on following accepted network protocols and applying common packet capture methods. This approach forms the basis for comprehending and evaluating network communication in the context of TCP.

Comprehensive packet data capture, including send rate, latency, throughput, and block size, followed the Linux system's successful installation of TCP and BBR. Within the network context, this dataset serves as the foundation for in-depth research and performance evaluation[3].

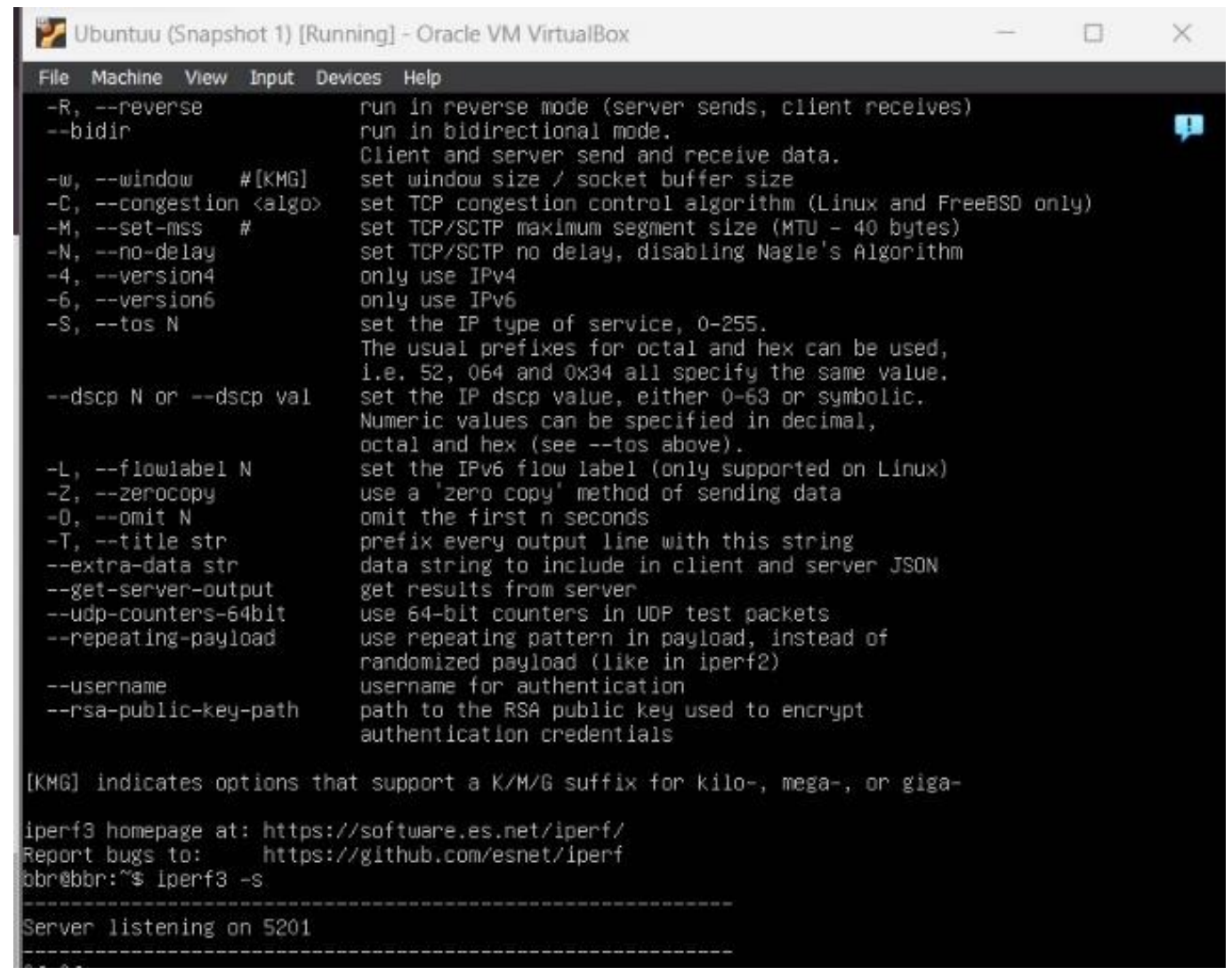

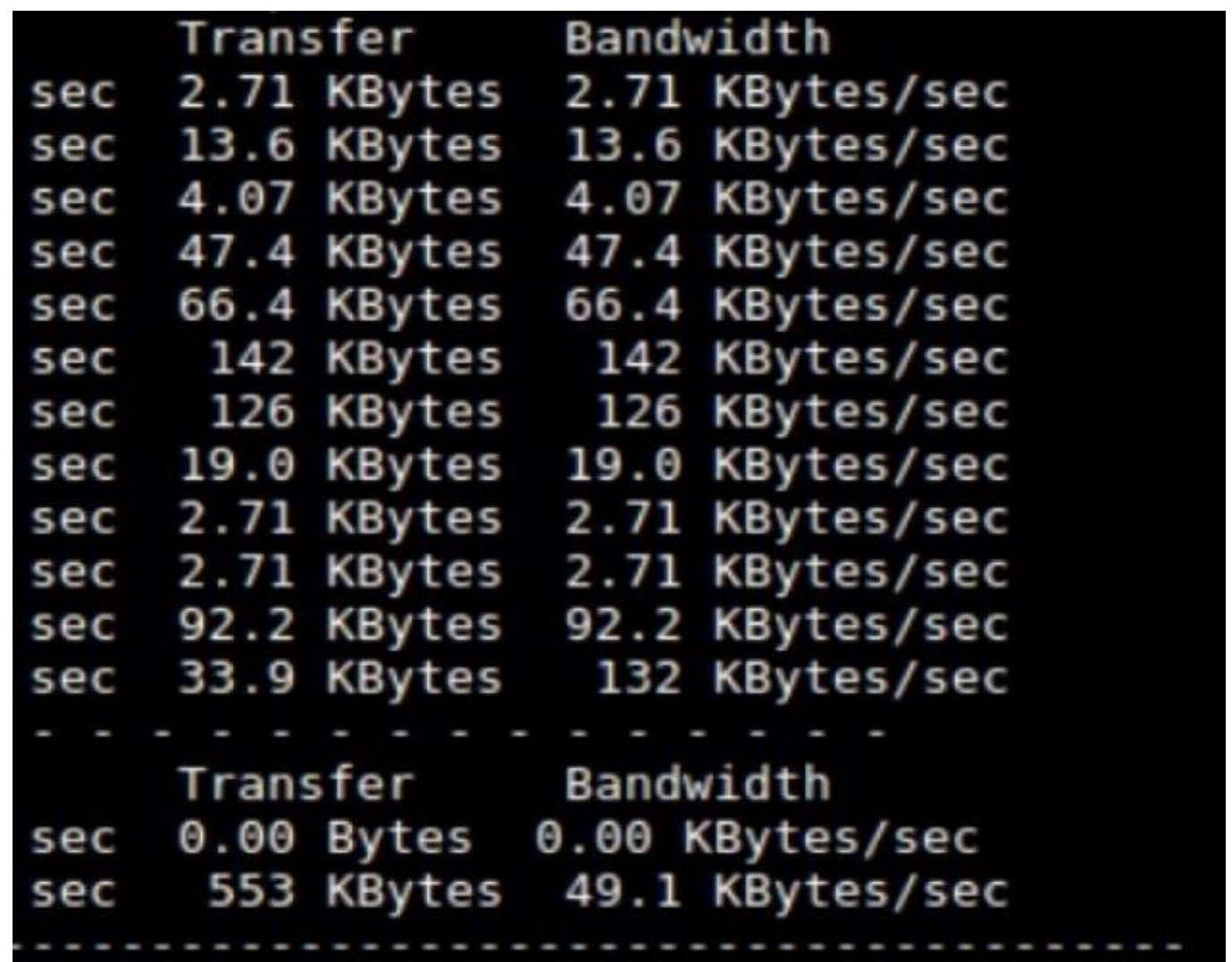

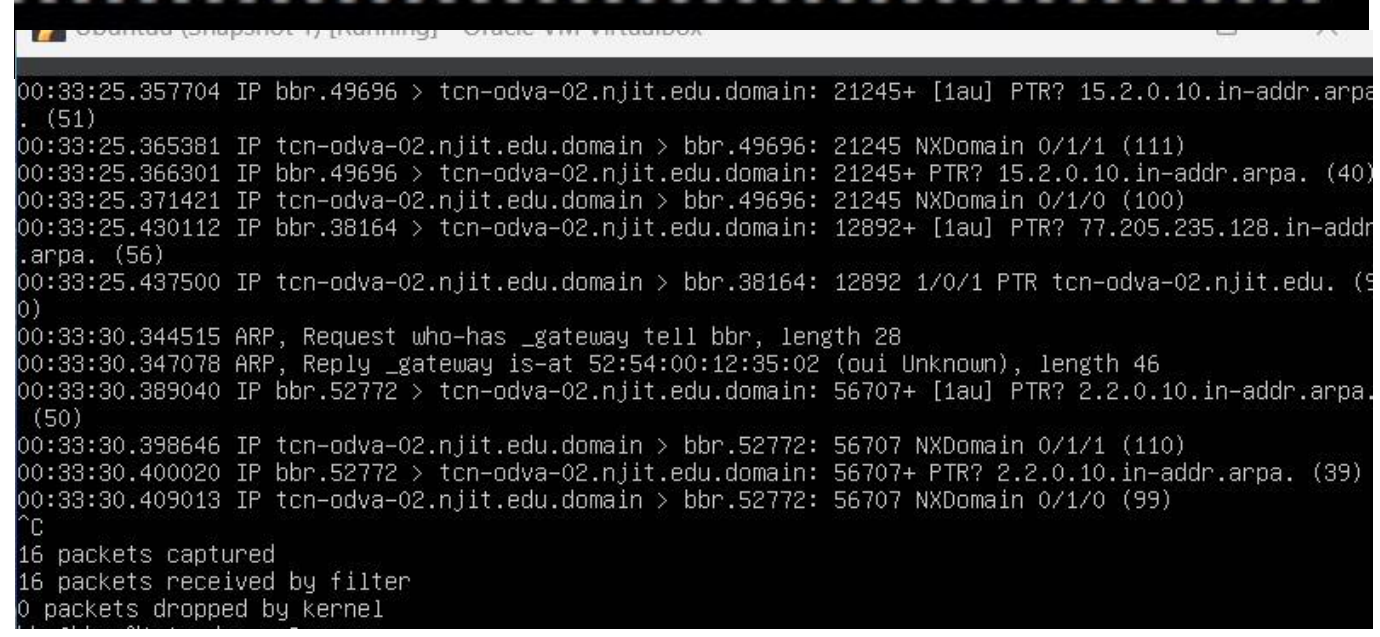

## 3. Experiments

### *3.1 Captured Data Analysis*

Based on the data captured, we examined the relationship between send rate and latency in a dataset used for machine learning model training. The y-axis shows the latency in seconds, and the x-axis shows the send rate. Each data point represents an individual measurement of latency at a particular send rate. The graph also shows that there is a general trend of increasing latency as the sending rate increases as shown in Figure (a). This means that it takes longer to send data as the number of data points being sent increases. The graph shows the relationship between the number of messages sent per second (send rate) and the number of messages successfully delivered per second (throughput). The graph shows that as the send rate increases, the actual throughput initially increases but eventually reaches the maximum throughput (160 messages per second) around 100 send rate. It highlights the relationship between send rate and successful delivery, indicating the potential limitations or bottlenecks that may arise at higher message loads as shown in Figure (b).

Further we analyzed the relationship between send rates and latency for the three buffer sizes (10, 50, and 100) shown in the graph. All three lines show that latency generally increases as the send rate increases. This is because the system has to handle more messages at higher rates, which can lead to congestion and delays. The buffer size also has an impact on latency.

For a given send rate, the larger the buffer size, the lower the latency. This is because a larger buffer can temporarily store more messages, smoothing out fluctuations in the send rate and reducing the chance of messages being dropped due to congestion. The difference in latency between the buffer sizes is more pronounced at higher send rates. This suggests that the impact of buffer size is more significant when the system is under heavy load as shown in Figure (c).

Lastly, we analyzed the relationship between send rates and throughput for the three buffers (10,50,100). All three lines showed that throughput initially increases as the send rate increases, but eventually reaches a maximum value and plateaus. This is because the system can efficiently handle messages up to a certain rate, but beyond that, it becomes overloaded and starts dropping messages, leading to a decrease in throughput. The larger the buffer size, the higher the maximum throughput achieved. This is because a larger buffer can temporarily store more messages, allowing the system to handle bursts of traffic more effectively and reducing the chance of message drops as shown in Figure (d).

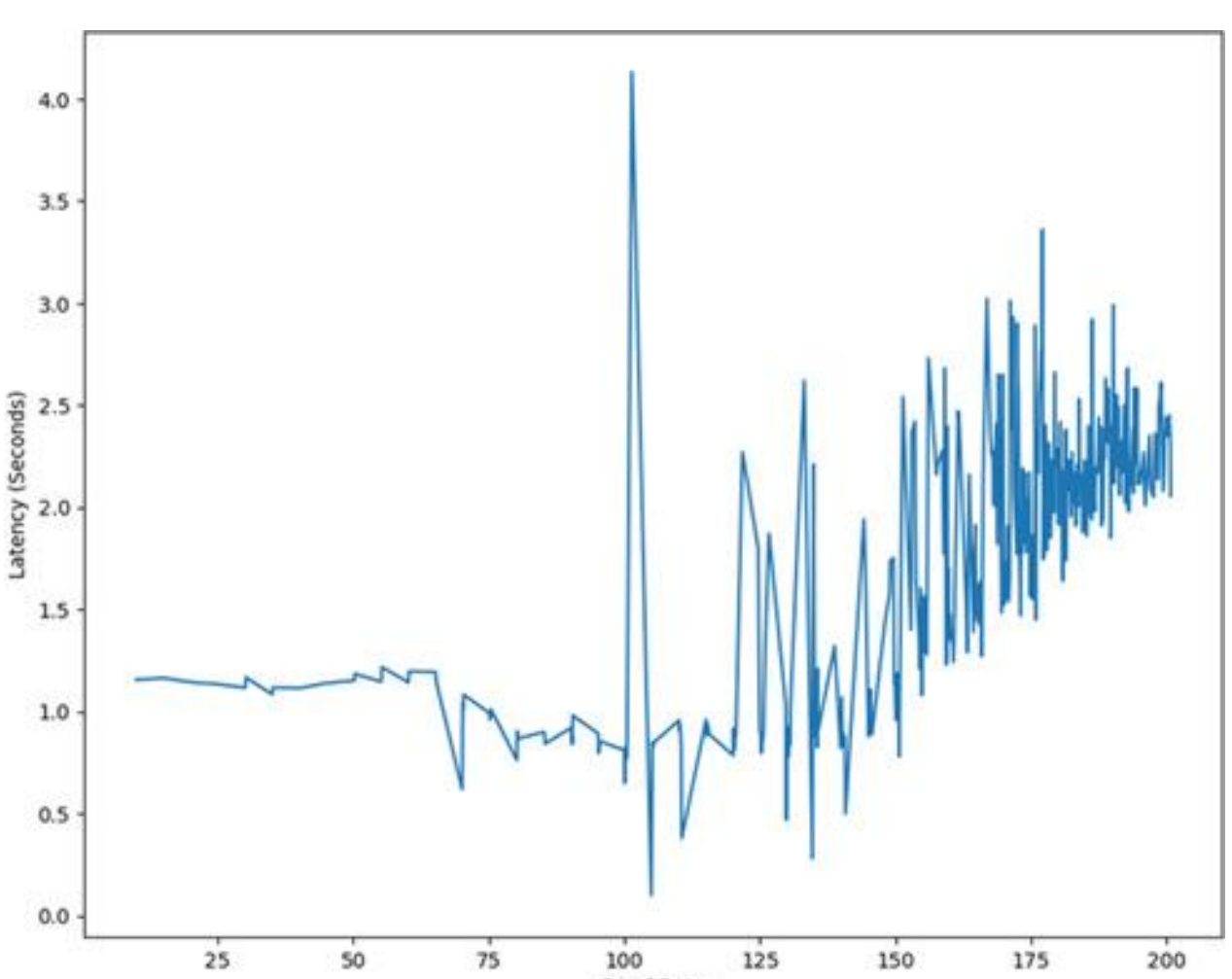

**Figure 3.1 (a) Average Latency**

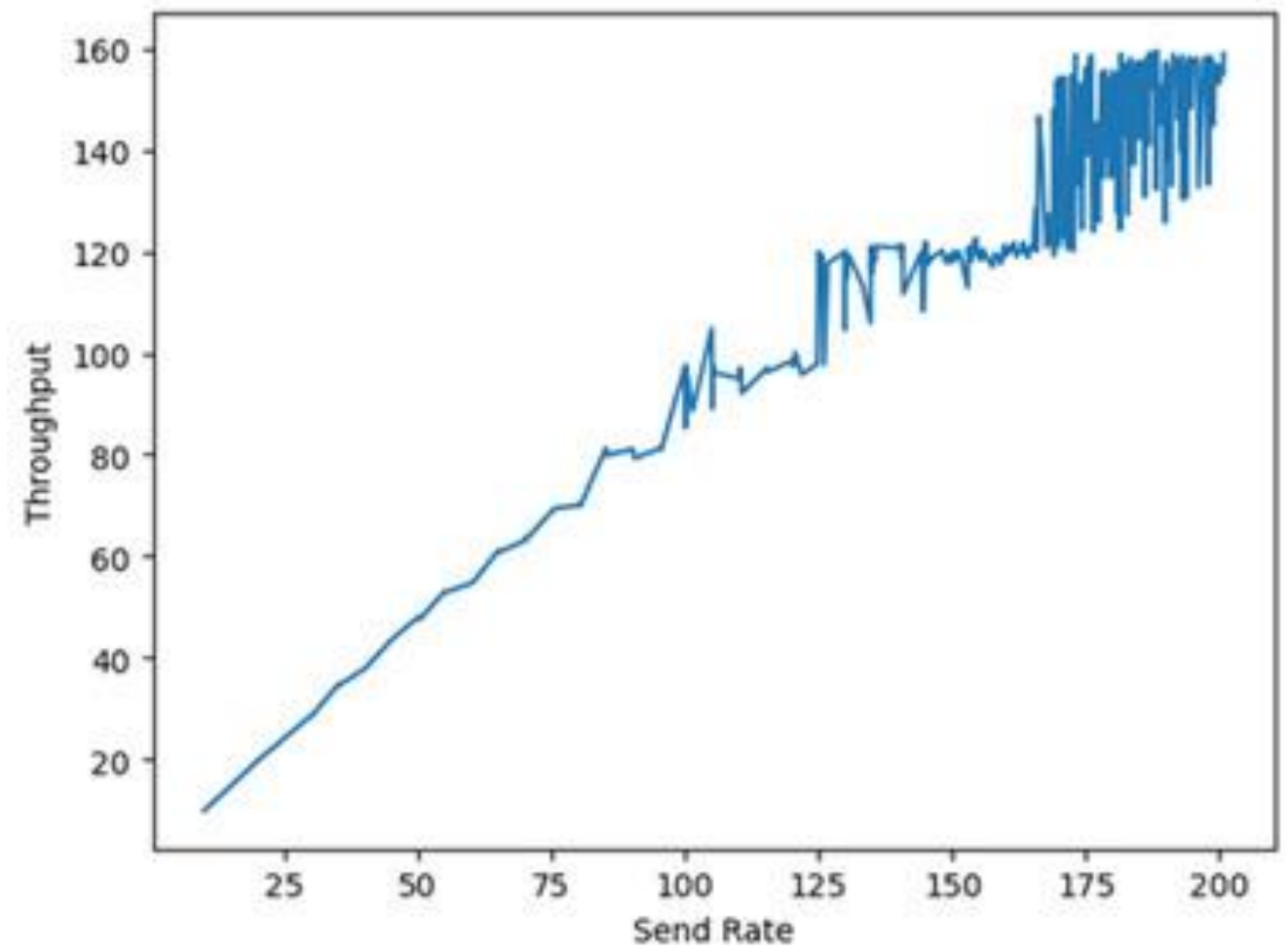

**Figure 3.1 (b) Average Throughput**

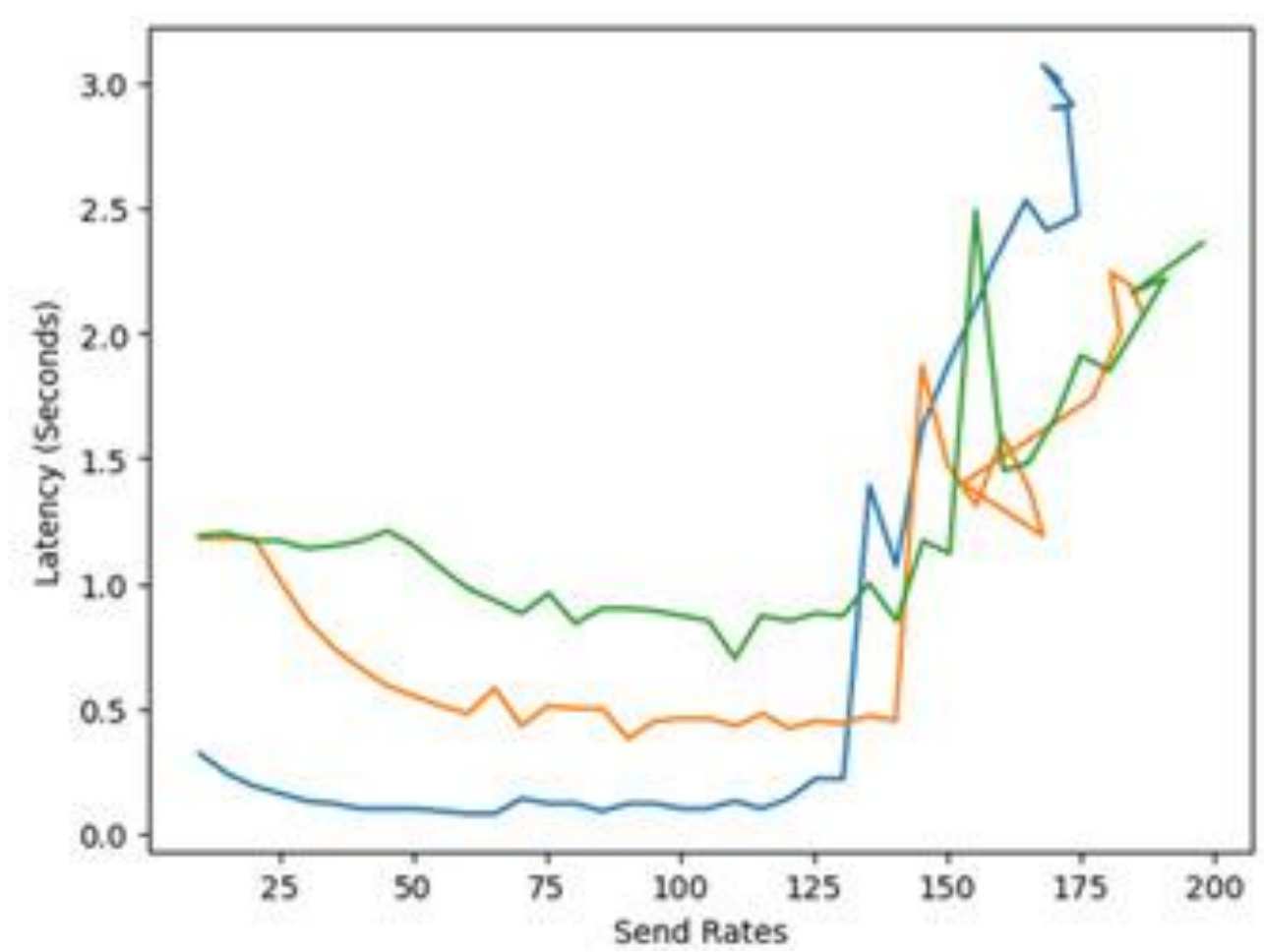

**Figure 3.1 (c) The effect of block size for latency**

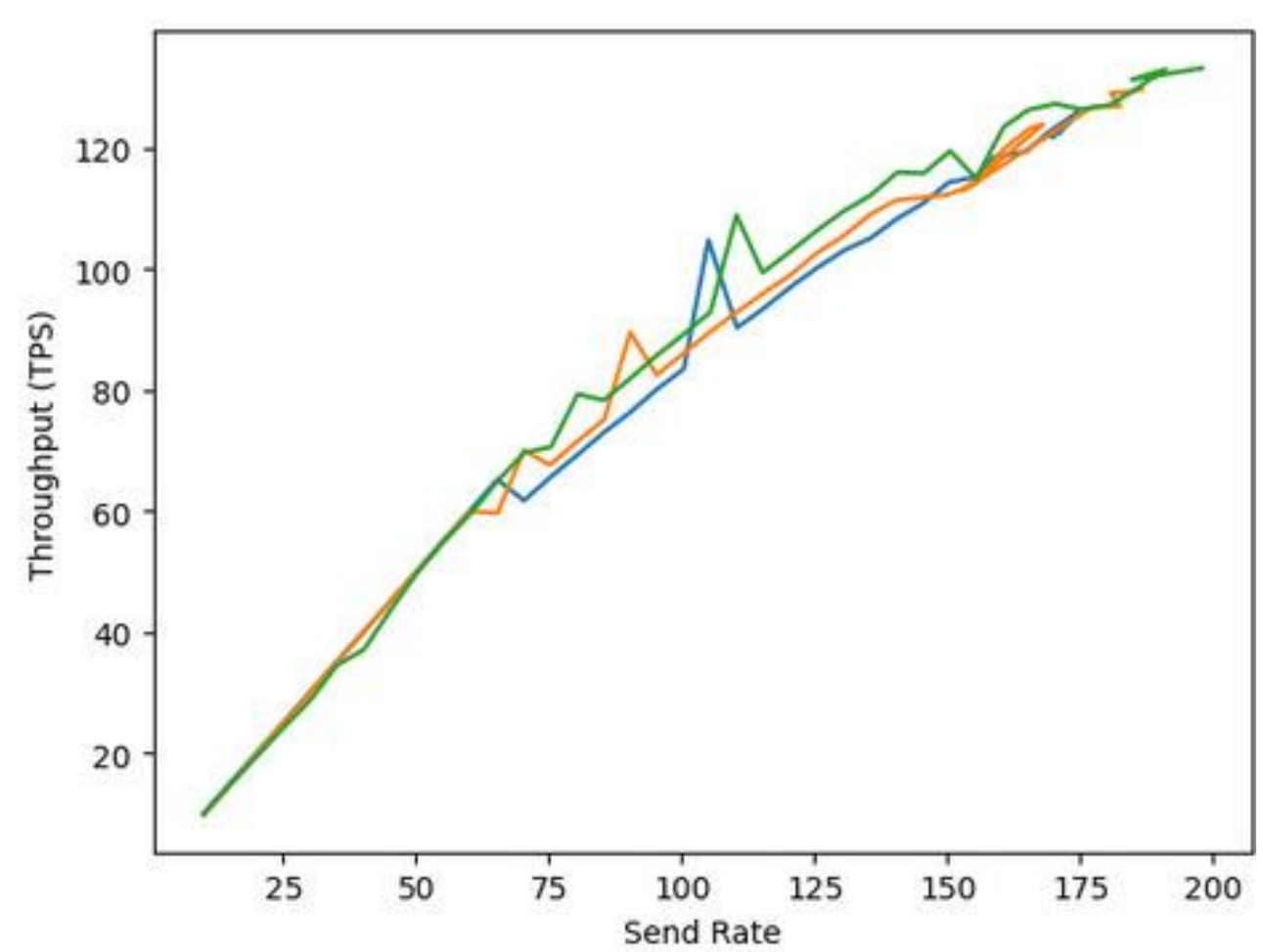

**Figure 3.2 (d) The effect of block size for throughput**

### *3.2 Machine Learning Algorithms*

We used a machine learning pipeline with common classifiers from the scikit-learn package to analyze and compare the effectiveness of several classification models in predicting latency classes based on characteristics such as block size and throughput. Several critical phases were involved in the implementation, including data preparation, model training, and performance evaluation.

To make the classification work easier, we first constructed a binary target variable, 'Latency Class,' by applying a threshold to the 'Avg Latency' characteristic. Instances with average latency values equal to or less than the threshold were branded 'Low,' while those with latency values more than the threshold were labeled 'High.'

We looked at three common classification algorithms: SVM, Decision Tree, and Multi-layer Perceptron (MLP) [4]. We created a classifier for each method and trained it on the training data as shown in Figure 3.2(a).

1. Support Vector Machine (SVM): We used the scikit-learn SVC class with the default hyperparameters.
2. For tree-based classification, the Decision Tree Classifier class was used.
3. MLP Classifier: A neural network classifier was instantiated using the MLP Classifier class.

```
Import necessary libraries
# Define a threshold to convert Avg Latency to classes
    threshold = 1.0
# Define features (X) and target variable (y)
    X = data[['Block Size', 'Throughput']]
    y = data['Latency Class']
# Split the data into training and testing sets
# Standardize the features
    scaler = StandardScaler()
# Initialize lists to store accuracy values
    svm_accuracies = []
    tree_accuracies = []
    mlp_accuracies = []
for _ in range(num_runs):
    svm_accuracy = np.mean(svm_cv_scores)
    tree_accuracy = np.mean(tree_cv_scores)
    mlp_accuracy = np.mean(mlp_cv_scores)
accuracy_df = pd.DataFrame
    'SVM': svm_accuracies,
    'Decision Tree': tree_accuracies,
    'MLP': mlp_accuracies
# Print mean accuracies
```

**Figure 3.2(a) Machine Learning Algorithm**

In networking, using a neural network classifier for BBR provides a sophisticated way to improve inter-protocol fairness. Because neural networks can record complicated relationships in data, the BBR algorithm can dynamically adapt to changing network circumstances. This flexibility improves fairness by optimizing resource allocation across protocols, minimizing biases, and encouraging equal performance across a wide range of contexts [5].

The model is trained using the Adam optimizer and the Mean Squared Error loss. For 1000 epochs, the training loop updates the model's parameters to minimize the loss. The algorithm assesses the predicted accuracy of the trained model using the same data and calculates the Mean Squared Error. Consequently, an improved regression model capable of forecasting BBR throughput with little error has been developed as shown in Figure 3.2(b).

**Figure 3.2(b) Neural Network Classifier Algorithm**

```
# Define the input and output variables
Input: X, a matrix of features
Output: y, a vector of target values
Class RegressionModel:
    Function _init_(input_size, hidden_size, output_size):
    # Define the linear layers and the activation function
# Define the forward pass
  Function forward(x):
    # Pass the input through the linear layers and the activation function
    Return x
model = RegressionModel(input_size, hidden_size, output_size)
# Define the loss function and the optimizer
criterion = nn.MSELoss()
optimizer = optim.Adam(model.parameters(), lr=0.01)
# Train the model
For epoch in range(num_epochs):
  optimizer.zero_grad()
  outputs = model(X)
  loss = criterion(outputs, y)
With torch.no_grad():
  model.eval()
  predicted_throughput = model(X).numpy()
mse = nn.MSELoss()
mse_value = mse(torch.tensor(predicted_throughput), y)
```

## 4. Evaluation

Machine learning models, particularly Support Vector Machine (SVM), Decision Tree, and Multi-layer Perceptron (MLP)[6], have given encouraging findings in terms of forecasting latency outcomes in the context of our inquiry into the BBR method.

The observed accuracies demonstrate the models' capacity to recognize intricate correlations, allowing the BBR algorithm to be optimized for various contexts. This, in turn, helps to more equal resource allocation among various protocols, reducing biases and encouraging enhanced inter-protocol fairness.

**Figure 4 (a) Machine Learning Model Accuracies**

```
Mean SVM Accuracy: 0.8589771543086172
Mean Decision Tree Accuracy: 0.8671086172344689
Mean MLP Accuracy: 0.8575364328657316
```

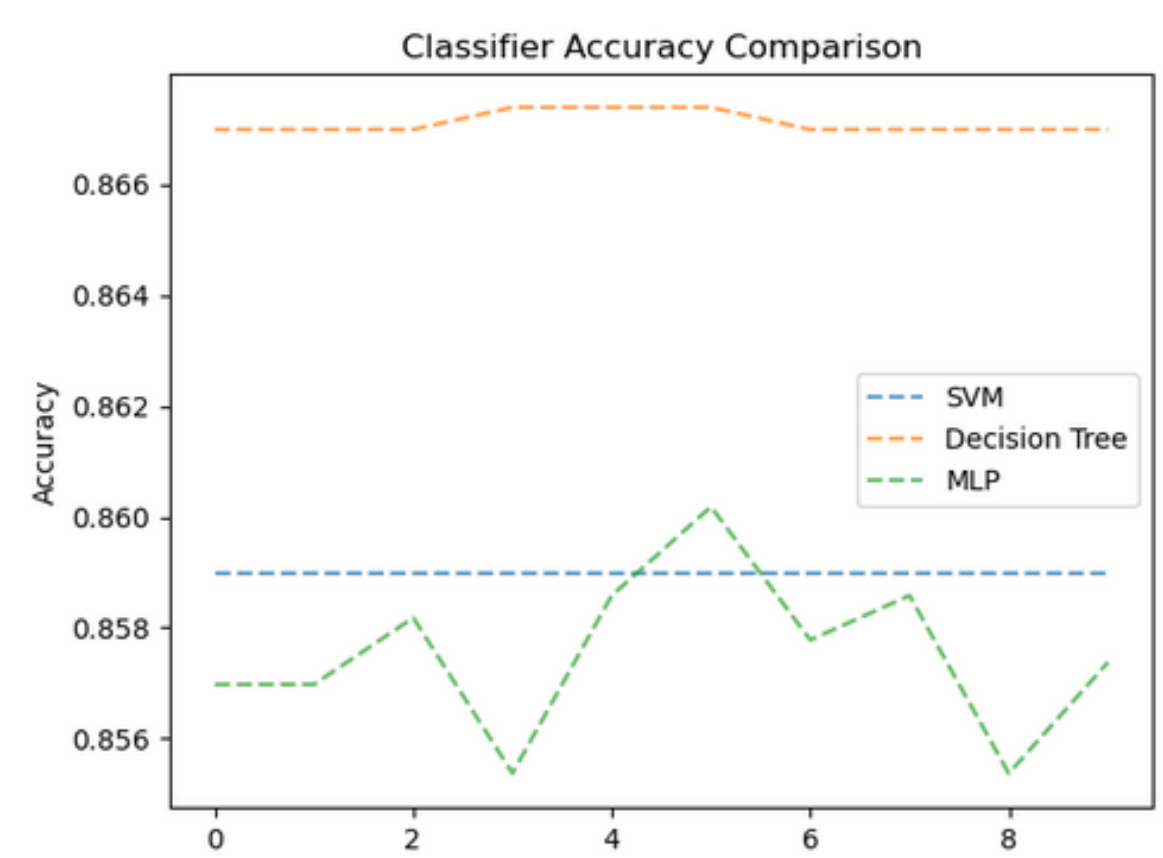

Furthermore, the resilience of these models, as demonstrated by the constancy of accuracies across numerous runs, improves their generalizability. This is crucial for assuring optimal BBR algorithm performance under a variety of settings, hence strengthening the models' effectiveness in promoting fairness across a wide range of network scenarios.

The neural network classifier's accuracy of 83.17% on the validation set indicates a strong capacity to predict latency levels in the BBR method. This precision is critical for optimizing BBR's decision-making process, which leads to improved inter-protocol fairness. The classifier's capacity to distinguish between low and high latency cases enables proactive modifications, resulting in a more responsive and flexible BBR implementation. Using machine learning, particularly neural network classifiers, is beneficial in addressing and minimizing latency fluctuations, resulting in a more equitable distribution of network resources.

```
Epoch 0, Loss: 9.533106803894043
Epoch 100, Loss: 0.48956266045570374
Epoch 200, Loss: 0.47000837326049805
Epoch 300, Loss: 0.5210153460502625
Epoch 400, Loss: 0.4501846432685852
Epoch 500, Loss: 0.46196249127388
Epoch 600, Loss: 0.45421722531318665
Epoch 700, Loss: 0.4534854292869568
Epoch 800, Loss: 0.3910413682460785
Epoch 900, Loss: 0.35858938097953796
Accuracy on Validation Set: 0.8317307692307693
```

**Figure 4 (b) Neural Network Accuracy**

## 5. Conclusion

In conclusion, the research highlights the pertinent issue of fairness in BBR, where varying network conditions pose challenges to equitable resource distribution. Employing machine learning models, particularly the neural network classifier, emerges as a promising solution to mitigate latency variations. The accuracy achieved on the validation set (83.17%) attests to the classifier's effectiveness in predicting latency levels, contributing to a more responsive and adaptive BBR algorithm. The utilization of neural network classifiers holds the key to enhancing inter-protocol fairness by providing a reliable means to preemptively adjust network parameters. One future scope lies in further refinement of the classifier to accommodate dynamic network scenarios and evolving protocols, ensuring a sustained improvement in inter-protocol fairness within BBR.